%% file: main.tex
\documentclass[sigconf]{acmart}
\acmConference[SE 2030]{International Workshop on Software Engineering in 2030}{November 2024}{Trondheim, Norway} 

\usepackage{xspace}
\usepackage{hyperref}

\title{Towards Evidence-Based Tech Hiring Pipelines }

\author{Chris Brown}
\orcid{0000-0002-6036-4733}
\email{dcbrown@vt.edu}
\affiliation{%
  \institution{Virginia Tech}
  \city{Blacksburg, VA}
  \country{USA}}

\author{Swanand Vaishampayan}
\orcid{0009-0009-0664-8493}
\email{swanandsv@vt.edu}
\affiliation{%
  \institution{Virginia Tech}
  \city{Blacksburg, VA}
  \country{USA}}

\begin{abstract}
    Software engineers are responsible for developing, maintaining, and innovating software. To hire software engineers, organizations employ a \textit{tech hiring pipeline}. This process typically consists of a series of steps to evaluate the extent to which applicants meet job requirements and can effectively contribute to a development team---such as resume screenings and technical interviews. However, research highlights substantial flaws with current tech hiring practices---such as bias from stress-inducing assessments. As the landscape of software engineering (SE) is dramatically changing, assessing the technical proficiency and abilities of software engineers is an increasingly crucial task to meet technological needs and demands. In this paper, we outline challenges in current hiring practices and present future directions to promote fair and evidence-based evaluations in tech hiring pipelines. Our vision aims to enhance outcomes for candidates and assessments for employers to enhance the workforce in the tech industry.
\end{abstract}

\input{alias.tex}

\begin{document}

\maketitle

\section{Introduction}

Software-based systems are increasingly impacting modern life across industries~\cite{andreessen2011software}. Despite recent innovations in the capabilities of large language models (LLMs) to automate software engineering (SE)-related tasks, human software engineers remain necessary~\cite{mastropaolo2024rise}. For instance, prior work demonstrates human expertise is necessary for analyzing requirements, designing, developing, and testing applications. To employ software engineers, organizations often leverage a \textit{tech hiring pipeline}~\cite{behroozi2020debugging}---or a multi-stage process to assess the fit of programmers for SE-related positions. Hiring pipelines provide many benefits in end-to-end hiring processes, including reduced time-to-hire, cheaper cost-per-hire, better quality hires, and increased retention~\cite{discovered}. Effective hiring evaluations are critical for stakeholders in tech hiring pipelines---namely, \textit{candidates}, individuals actively seeking SE jobs, and \textit{employers}, individuals assessing candidates to hire tech workers.

Hiring pipelines include a variety of recruiting related activities that impact candidates and employers---ranging from formulating job descriptions to extending and negotiating offers with top candidates~\cite{discovered}. They typically incorporate multiple levels of evaluation, including \textit{screening} processes to identify suitable candidates from the applicant pool and \textit{assessments} to evaluate candidates' skills and fit~\cite{airswift}. Tech hiring pipelines in particular often consist of two practices to evaluate the technical abilities of candidates: (1) \textit{resume matching}---screening candidates' resumes based on job description requirements; and (2) \textit{technical interviews}---assessing candidates' abilities through problem-solving and communication by completing programming challenges in front of employers.

While these practices are widely used in modern tech hiring pipelines~\cite{navigating}, they fail to adequately evaluate the technical proficiency of SE candidates. Prior work highlights numerous issues in current tech hiring pipelines that can inhibit candidate evaluation processes. For instance, resume matching in tech hiring can be biased~\cite{houser2019can} and misrepresent applicants' skills~\cite{fritzsch2021resume}.  Moreover, tech interviews are viewed as frustrating~\cite{mahnaz2019hiring}, stressful~\cite{behroozi2020does}, biased~\cite{lunn2022need}, and irrelevant~\cite{behroozi2020debugging}, in addition to being costly for employers~\cite{HiddenCost}. 

These challenges lead to issues such as false positive (selecting candidates who are ill-equipped or bad fits for development teams) and false negative (rejecting candidates who would be successful) hires in the software industry~\cite{awful}.  Further, the increasing usage of artificial intelligence (AI) in hiring pipelines~\cite{li2021algorithmic}---\ie resume parsing~\cite{JobScan} and job interviews~\cite{hickman2024developing}---can further complicate evaluations of candidates' technical abilities. For instance, prior work shows candidates distrust AI-based resume parsers used in tech hiring~\cite{swanand2022arp} and attempt to deceive AI screening systems~\cite{whitefont}. In addition, LLMs in technical interview contexts have been shown to provide inaccurate feedback on candidates' performance~\cite{swanand2025chase}.

Despite the fact that tech hiring has slowed due to ``economic uncertainty'' from the COVID-19 pandemic and the rise of generative AI~\cite{computerworld}, sources expect the tech job market to grow rapidly by 2030~\cite{nn}---with the World Economic Forum Future Jobs Report predicting at least 170 million new technology, data, and AI-related roles will be created in the next five years~\cite{wef}. Thus, efficient and effective evaluations of SE candidates is crucial for supporting the future systems that impact human behavior and well-being. To improve hiring processes for candidates and employers, prior work posits \textit{evidence-based hiring} to use data-driven approaches for selecting job candidates~\cite{evidence_hiring}. However, tech hiring pipelines primarily rely on intuition-based hiring~\cite{miles2014recruitment}---lacking evidence to effectively evaluate candidates' skills~\cite{mahnaz2019hiring,swanand2025chase}. To this end, our vision paper outlines challenges and future directions to integrate contextual, experiential, and research-based evidence into tech hiring pipelines---particularly resume matching and tech interviews---to improve assessments and accurately demonstrate the technical abilities of candidates pursuing SE-related roles in the tech industry.

\section{Background}

This section provides a broad overview of evidence-based hiring, AI usage in hiring contexts, and current tech hiring practices. 

\subsection{Evidence-Based Hiring}

Prior work suggests resumes and interviews do not provide sufficient information to make effective hiring decisions~\cite{tricarico2024evidence}. Evidence-based hiring aims to use data-driven approaches to select the best candidates for open positions, improving hiring pipelines for candidates and employers~\cite{evidence_hiring}. Prior work introduces a variety of methods and resources to promote evidence-based hiring practices~\cite{sackett2022revisiting,carless2009psychological,rousseau2011becoming}. Research also demonstrates the effectiveness of evidence-based hiring in various domains, such as education~\cite{sekaquaptewa2019evidence,cohen2019evidence,cohen2011staffing} and nursing~\cite{mcdonald2012review}. However, there is a notable gap in  support for evidence-based hiring in the tech industry. For instance, Behroozi et al. highlight developers' complaints about the lack of evidence in tech hiring practices~\cite{mahnaz2019hiring}. We aim to fill this gap exploring ways to incorporate evidence-based hiring in tech hiring pipelines.

\begin{figure}
    \centering
    \includegraphics[width=0.4\textwidth]{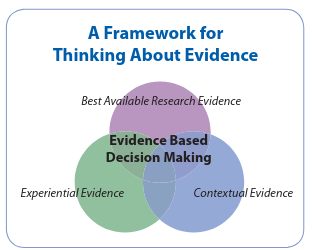}
    \caption{Evidence-based Decision Making Framework, as portrayed by~\cite{CDC2}.}
    \label{fig:evidence}
\end{figure}

Evidence-based hiring is a subset of evidence-based decision making, which is grounded in using the best available data to enhance decisions~\cite{baba2012evidence}. Research shows evidence-based decision making is critical in a variety of fields, such as crime prevention~\cite{bania2018evidence,CDC}, public health~\cite{wahid2019evidence}, maternal and child health~\cite{powis2022shifting}, and management~\cite{baba2012evidence}. Researchers have explored methods to promote evidence-based SE---improving behavior and decision making in software development contexts based on research evidence~\cite{kitchenham2004evidence,dyba2005evidence}. Prior work suggests evidence from diversified sources---such as specific contexts, experiences, and research---are vital for effective decision making~\cite{APA,satterfield2009toward,CDC}. We leverage this framework, presented in Figure~\ref{fig:evidence}, to outline future research directions to promote evidence-based decision making tech hiring pipelines.

\subsection{AI-Based Hiring}

Recently, there is an increasing use of generative AI in hiring pipelines. Studies show incorporating AI in hiring pipelines can streamline hiring and talent sourcing tasks~\cite{koumpan12024revolutionizing,gan2024application}. For example, AI-based systems have been explored in automating a variety of hiring tasks, including generating job descriptions~\cite{walker2025leveraging}, cover letters~\cite{bagowrite}, reference letters~\cite{wan2023kelly}, and job offer/rejection emails~\cite{an2024measuring}. For instance, sources suggests over 99\% of Fortune 500 companies utilize AI-based applicant tracking systems---platforms to automatically parse and rank applicants based on resumes~\cite{JobScan}. Research also shows LLMs are effective for resume matching tasks~\cite{li2020competence}---demonstrating comparable performance to humans~\cite{swanand2025acl}. Moreover, AI is increasingly used to administer~\cite{liu2023speech} and score~\cite{hickman2024developing} job interviews. In SE hiring contexts, AI is increasingly used to help candidates prepare for tech interviews. For instance, studies explore the capabilities of LLMs in mock tech interview settings~\cite{chun2024study,mejias2023equity}.

However, algorithmic hiring also comes with challenges. For instance, prior work shows recruiters perceive AI tools as inaccurate and complex~\cite{li2021algorithmic}. Candidates are also wary of incorporating AI in hiring pipelines~\cite{pew}. For example, prior work shows candidates find AI-based resume parsers used in tech hiring lack transparency and are untrustworthy---significantly preferring human raters~\cite{vaishampayan2023procedural}. Research also show candidates distrust feedback from LLMs in mock interviews~\cite{swanand2025chase}. In addition, prior work shows AI-based resume screening can incorporate bias against marginalized populations~\cite{an2024large,yarger2020algorithmic}, such as ethnicity/race~\cite{armstrong2024silicone}, disability~\cite{glazko2024identifying}, and gender~\cite{meyer2018amazon}. In this work, we aim to explore methods to incorporate evidence demonstrating the technical abilities of candidates into modern AI-based hiring pipelines.

\subsection{Tech Hiring Pipelines}

While there are many aspects of tech hiring pipelines (\ie job description formulation, offer negotiation, etc.), the scope of this work focuses on two hiring evaluation processes within tech hiring pipelines---\textit{resume matching} and \textit{technical interviews}.

\subsubsection{Resume Matching}

Typically the first evaluative stage of the tech hiring pipeline is resume matching. Resume matching evaluates the extent to which the content of applicants' submitted resumes---documents presenting a summary of their professional goals, abilities, and background~\cite{mcdowell1987perceptions}---matches the position requirements outlined in job descriptions~\cite{swanand2025acl}. As the number of resumes submitted to job openings rises, manual resume matching techniques are infeasible in modern hiring contexts~\cite{torres2million}. Thus, AI tools are increasingly used to support resume matching tasks---including parsing resume content, extracting keywords relevant to job descriptions, and ranking resumes to compare against other candidates applying for the same position~\cite{rchilli-keyword}.

\subsubsection{Technical Interviews}

Candidates who pass resume matching advance to the next stage of tech hiring pipelines---typically one or more rounds of technical interviews. This specialized interview process requires candidates to solve data structures and algorithms-focused challenges by writing code or pseudocode on a whiteboard or other non-programming environment while simultaneously communicating their problem-solving approach to employers via think-aloud~\cite{whiteboard}. This process is critical in tech hiring, providing insights on technical and communication skills~\cite{ford2017tech}. However, prior work suggests this process is ineffective for evaluating candidates~\cite{mahnaz2019hiring}. For instance, problem-solving in front of employers can induce performance anxiety---hindering candidates' performance in interviews~\cite{behroozi2020does}.

\section{Roadmap to Evidence-Based Tech Hiring}

To promote evidence-based decision making in tech hiring pipelines, we provide future research directions based on contextual, experiential, and research-based evidence (see Figure~\ref{fig:evidence})~\cite{CDC,CDC2}.

\subsection{Contextual Evidence}

Contextual evidence refers to information about whether a process or strategy ``fits'' within a certain decision-making context~\cite{CDC}. For example, collecting evidence from local data sources within communities (\ie schools, census, economic, etc.) to determine whether violence prevention strategies are feasible, useful, and likely to be accepted in particular communities~\cite{CDC-context}. In tech hiring pipelines, we posit contextual evidence as data to demonstrate whether a candidate has the capability to effectively contribute to a specific SE context---such as a development team.

\subsubsection{Current Challenges}

Current tech hiring processes fail to accurately assess the abilities of candidates within the hiring context. Resume matching relies on the content of candidates' resumes to assess skills---however, resumes are not ideal for providing contextual evidence. For example, the length of resumes is an important factor in hiring decisions, with sources recommending resume length to be no more than two pages maximum~\cite{blackburn2001one}. While this lengths makes resumes easier to review and longer documents are ``at risk of being disregarded''~\cite{foundry}, it limits the content which can be put into the resume. For instance, candidates often omit contextual information from their resume to fit details within one-page guidelines~\cite{ross2005resume}. Finally, candidates may choose to personalize and customize resume documents to incorporate additional context beyond conventional resume templates---for instance, two-column resumes. However, specially formatted resumes may be unreadable by automated AI-based resume parsing systems~\cite{indeed}. 

Technical interview settings---writing code  in front of observers---do not reflect industry and computing education contexts~\cite{lunn2022need}, and thus fail to provide contextual evidence of candidates' skills. Developers view tech interviews as irrelevant to real-world SE work~\cite{mahnaz2019hiring}. A major concern is the questions asked in technical interviews do not reflect relevant SE work. Prior work shows the primary frustration with current interview practices is the lack of relevance, with problem-solving not grounded in real-world constraints~\cite{mahnaz2019hiring}. For instance, Max Howell---a developer who created the Homebrew package manager\footnote{\url{https://brew.sh/}} utilized by over 90\% of software engineers at Google---was unable to obtain a position at Google due to unsuccessfully inverting a binary tree on a whiteboard~\cite{max}. Candidates spend substantial time preparing for tech interviews~\cite{cui2024much}, causing anxiety and stress~\cite{hall2018effects,bell2023understanding}. In addition, candidates often focus on gaining knowledge to succeed in interviews instead of gaining skills to successfully contribute to software development projects~\cite{LeetCoders}. Ultimately, tech interviewing has become a ``question lottery'' based on ``luck and chance'' rather than assessing the skills and abilities of candidates~\cite{Yalkabov2016}. To this end, novel solutions are needed to resume and interviewing evaluation processes to obtain contextual evidence outlining candidates' ability to contribute to software teams and products.

\subsubsection{Future Directions}

\paragraph{Longer Resumes}

To provide contextual evidence in resume matching, one consideration is to promote longer resume documents. As humans are less involved in resume matching, longer documents can include more context on candidates' experiences---providing additional insights on their ability to contribute to the development team. Further, despite the fact one page is recommended for resumes~\cite{foundry}, prior work shows recruiters for entry-level positions in recreation and leisure services prefer longer resumes---opting for two to four pages in length and for resume length to be commiserate with the amount of information~\cite{ross2005resume}. Research shows LLMs are capable of parsing and understanding long documents~\cite{zou2024docbench}, including demonstrating the ability to effectively navigate and modify large code bases~\cite{shah2025students}. Thus, researchers and practitioners can explore the effects and capabilities of using AI-based systems to extract relevant information from longer documents outlining the skills and capabilities of candidates.

\paragraph{Move Beyond Resumes}

Most employers require prospective candidates to submit resumes describing their qualifications for open job positions---however, this job application tradition dates back to the 1400s~\cite{sprocket}. Resumes are also difficult for candidates to maintain. For instance, updating resumes is time-consuming, especially if candidates tailor resumes to specific job postings~\cite{coconut}. Future work can explore more modern techniques to ascertain preliminary data on candidates' skills. For example, online job profiles on websites such as LinkedIn\footnote{\url{https://www.linkedin.com/}} and Handshake\footnote{\url{https://joinhandshake.com/}} allow users to showcase their professional experiences, certifications, and skills to potential employers. Further, research show online job profiles are useful for personalized tech job recommendations~\cite{kumalasari2019recommendation}, matching candidates to job openings \cite{koch2018impact}, and helping users advance in their careers~\cite{pan2017understanding,stone2015digital}. We posit incorporating online job profiles into hiring pipeline evaluations can provide better contextual evidence of candidates' abilities to contribute to development teams compared to analyzing resumes alone. 

\paragraph{Relevant Tech Interview Questions} Prior work has suggested methods to improve the context of technical interviews. For example, some companies leverage take-home assessments---where candidates are tasked with completing a coding problem within a given time frame~\cite{assess}. This approach can provide more insights on candidates' problem-solving abilities in more relevant contexts (\ie using their own tools and workflows). Other project-based assessments can also better evaluate candidates real-world contexts~\cite{byteboard}. While candidates may find these approaches preferable~\cite{HiringWithoutWhiteboards}, there is limited empirical research exploring the effectiveness of alternative hiring approaches in the tech industry. Further, irrelevant questions also lead to inconsistent evaluations of candidates---assessing their test-taking abilities instead of their problem-solving skills~\cite{ncsu-sector}. Research can also explore mechanisms to generate technical interview questions beyond typical LeetCode-style challenges consisting of data structures and algorithms-based questions currently used in interviews. For instance, prior work explores scenario-based questions---questions presenting hypothetical situations to which candidates respond---to enhance computing education~\cite{kerven2017scenario,zitouniatis2023teaching}, job interviews~\cite{zhang2019telling}, and on-the-job training~\cite{chine2022scenario} in different contexts. Similar approaches can be used to investigate how candidates would perform given specific constraints and real-world scenarios within software development contexts.

\paragraph{Contextualize Tech Interview Settings} Another challenge is technical interview environments do not reflect real-world software development work. For example, tech interviews typically require candidates to solve coding challenges in a non-programming environment, such as a whiteboard or using online text processing software such as Google Docs\footnote{\url{https://docs.google.com/}} or CollabEdit\footnote{\url{http://collabedit.com/}}. Research shows developers desire to use IDEs and code editors in technical interviews~\cite{mahnaz2019hiring}, incorporating helpful features such as code complete, syntax highlighting, compilers, etc. Recently, companies have explored this through allowing candidates to live screen share during coding exercises during technical interviews~\cite{swhelan}---which can provide insight on how familiar candidates are with various tools and technologies. With the rise of generative AI in software development~\cite{abrahao2025software}, sources have posited allowing the use of AI tools during interviews to observe how candidates use them to complete programming tasks~\cite{swe}. In addition, research has proposed additional techniques to enhance tech interview settings, such as pair programming between interviewers and interviewees~\cite{mahnaz2022async}. This would allow employers to gain insights on candidates' programming skills and their ability to collaborate with another developer in a development context. Implementing and assessing these approaches can provide insights on evaluating the suitability of candidates in relevant contexts in which tech hiring decisions are made.

\subsection{Experiential Evidence}

Experiential evidence refers to information based on the expertise and skills of individuals collected over a period of time~\cite{CDC-experience}. For instance, the clinical experiences of healthcare professionals (\ie interactions with patients) is critical for evidence-based decision making in maternal and child health~\cite{powis2022shifting}. In this context, we outline challenges and propose future directions to obtain experiential evidence accurately reflecting the expertise of candidates pursuing SE-related positions in the tech industry.

\subsubsection{Current Challenges} Currently, evaluation process in tech hiring pipelines fail to accurately evaluate the experiences of candidates. For instance, resume matching analyzes the extent to which candidates are qualified for positions. However, there are several limitations to this approach. For instance, most resume matching systems ascertain technical skills using keyword-based approaches~\cite{rchilli-keyword}. However, resumes missing the exact desired keywords are automatically filtered out even if candidates do have relevant experience. For example, not including a specific programming language on resumes will exclude candidates from consideration for positions---despite the fact programmers are often adept and capable of quickly learning new programming languages~\cite{shrestha2018towards}. Additionally, candidates can mislead employers based on the content of their resumes. For instance, the job search website Indeed\footnote{\url{https://www.indeed.com/about}} suggests about 40\% of candidates lie on their resumes~\cite{indeedlying}. Fritzsch et al. also show developers often overemphasize trending technologies on their resumes~\cite{fritzsch2021resume,fritzsch2023resist}. Moreover, recent trends such as ``white fonting'' resumes---adding specific keywords to resumes in a white colored font so they are visible to automated parsing systems but not to human recruiters---complicate assessments of the actual expertise of programmers~\cite{whitefont,cnbc}. For example, tools such as Inject My PDF inject invisible prompts into resumes to mislead and take advantage of AI-based resume screening systems~\cite{greshake2023not}.\footnote{\url{https://kai-greshake.de/posts/inject-my-pdf/}}

Technical interviews also fail to provide evidence on the experiences of candidates. For instance, some candidates attempt to game the current interview process by extensively studying LeetCode problems to be successful in interviews, but may not have the actual expertise~\cite{LeetCoders}. Research also shows technical interviews are biased towards candidates with more time and resources to prepare~\cite{mahnaz2019hiring}. Moreover, the current tech interview process fails to adequately represent the experiences of qualified candidates. For example, studies show a mismatch in technical interview expectations between candidates and employers~\cite{ford2017tech}. The experience and temperament of interviewers can also impact assessments~\cite{behroozi2020debugging}. Finally, research shows technical interviews induce performance anxiety due to the performative nature of problem-solving in front of an audience of employers~\cite{ncsu-sector}. In addition, time pressure can add further anxiety to interview settings~\cite{parnin2020does}. This anxiety has been shown to significantly decreases candidates' performance during tech interviews~\cite{behroozi2020does}, providing inaccurate assessments of their skills and expertise. In addition to the previously mentioned directions, we provide avenues of future work to enhance the ability to collect experiential evidence in tech hiring pipelines.

\subsubsection{Future Directions}

\paragraph{Experienced-Based Resume Screenings} Resume matching primarily relies on searching for keywords in resumes to find suitable candidates based on job description requirements. However, this method can be ineffective as different words can have the same semantic meaning (\ie ``JavaScript'' and ``JS'') and candidates may have relevant experiences not explicitly worded the correct way~\cite{vaishampayan2023procedural}. To mitigate this, research can explore methods to enhance the content of resumes with evidence from external data sources. For instance, Vaishampayan and colleagues introduced \texttt{GitMeter}, a novel tool for GitHub-supported resume matching that extracts links to candidates' GitHub\footnote{\url{https://github.com}} profiles from resumes and automatically provides insights on their technical abilities based on analyzing code contributions to public open-source software projects~\cite{swanand2025saner}. Similar approaches can be used to gain insights on candidates' experiences and skills from other resources, such as LinkedIn, Stack Overflow,\footnote{\url{https://stackoverflow.com}} GitLab,\footnote{\url{https://about.gitlab.com}} and other online programming communities. 

\paragraph{Other Sources of Experiential Evidence} Employers can also use other methods to gain evidence on candidates' abilities and skills in tech hiring pipelines. For example, letters of recommendation---or a document written on behalf of a candidate to outline their skills, qualifications, and experiences~\cite{indeed-letter}---can provide insights into candidates' skills and predict employment performance~\cite{aamodt1993predicting}. However, sources suggest recommendation letters are rarely required and under-utilized in tech hiring contexts~\cite{reference-cscq,abel2020value}. Similarly, tech workers specifically report that their professional references are rarely checked during hiring processes~\cite{references}. These methods can gain insights from individuals familiar with candidates beyond the content of their resume, providing evidence of their experiences and ability to effectively contribute to software projects.

\paragraph{Reduce Anxiety in Tech Interview Settings} Performance anxiety can inhibit candidates problem-solving abilities during tech interviews~\cite{behroozi2020does}---inadequately assessing the experiences and skills of candidates. For example, candidates' experiences highlight the challenges of being watched by employers during tech interviews led to poor performance caused by brain freezing~\cite{freeze}, severe anxiety and nervousness~\cite{nervous}, dizzyness and nausea~\cite{dizzy}, and ``my brain always turns to mush''~\cite{mush}. To mitigate these effects, research shows removing the performative aspects of tech interviews can enhance tech interview performance. For instance, Behroozi et al. show that candidates perform significantly better in private settings without an interviewer present than when being watched~\cite{behroozi2020does}. Research also demonstrates asynchronous technical interviews---where candidates submit recorded screencasts solving coding challenges---significantly improved candidates technical problem-solving and communication skills during interview assessments~\cite{mahnaz2022async}. Future work can explore additional techniques to reduce anxiety in tech interview settings, such as removing time limits~\cite{onwuegbuzie1995effect}, providing feedback~\cite{lu2024helping}, and other approaches to reduce social anxiety and negative self-thoughts~\cite{feiler2016role}.

Many employers also incorporate multiple types of interviews in hiring pipelines, such as phone screenings and behavioral interviews~\cite{bell2023understanding}. These qualitative assessments allow employers to gain additional insights on candidates' experiences. However, in tech hiring pipelines these assessments are often combined with technical interviews---conflating soft skills and technical skills evaluations~\cite{mahnaz2022async}. Incorporating additional stages in tech hiring pipelines can reduce tech interview anxiety and provide a greater understanding of the backgrounds of candidates, providing insights into their expertise and qualifications for specific roles.

\paragraph{Interviewer Perspectives} Interviewer experiences are also critical in hiring decisions. For example, research shows negative interactions with interviewers (\ie inexperienced, rude, etc.) can impact candidates' performance during technical interviews. Alternatively, positive interactions can enhance tech interview outcomes~\cite{behroozi2020debugging}. Thus, incorporating guidelines for interviewer behavior during interviews can provide more accurate assessments of candidates' abilities. There is extensive research exploring how candidates prepare for technical interviews (\ie~\cite{cui2024much,swanand2025chase,bell2023understanding,lunn2022need,hall2018effects})---however, there is limited work exploring how interviewers are trained and prepared for tech interviews. Moreover, evaluation and scoring can vary between interviewers, and the lack of transparency in hiring criteria is a source of frustration for developers~\cite{mahnaz2019hiring}. Specific rubrics and grading criteria can be used to provide more fair and consistent assessments of candidates' technical abilities~\cite{awful}, mitigating bias and reliance on intuition-based hiring decisions~\cite{miles2014recruitment}. Research can explore devising training resources and evaluation criteria to enhance the expertise of interviewers, providing more accurate insights into the experiences of candidates. 

\subsection{Research-Based Evidence}

Research-based evidence focuses on collecting information to support decision-making contexts from the best available literature and data derived from experiments~\cite{CDC-research}. For example, research based on clinical trials is critical to support evidence-based medicine~\cite{davidoff1995evidence}. To enhance tech hiring pipelines, we suggest areas to incorporate and improve research-based evidence to enhance hiring assessments for candidates and employers in the tech industry.

\subsubsection{Current Challenges}

Current tech hiring evaluations are not based on research evidence. Despite the fact research findings demonstrate that AI-based resume assessments face numerous issues---such as bias~\cite{meyer2018amazon,yarger2020algorithmic},they are increasingly used for resume matching in hiring pipelines~\cite{JobScan}. Similarly, research shows technical interviews are biased~\cite{lunn2021impact} and biased~\cite{mahnaz2019hiring}, yet they are the primary method for assessing the technical abilities of candidates~\cite{kaatz}. Here, we outline research directions to incorporate research-based evidence in tech hiring pipelines.

\subsubsection{Future Directions}

\paragraph{Responsible AI in Hiring Pipelines}

Research shows AI and machine learning models can exhibit bias in hiring tasks~\cite{an2024large}, including resume screening~\cite{meyer2018amazon} and interview scoring~\cite{booth2021bias}. In tech hiring pipelines, prior work demonstrates candidates distrust of AI-based resume parsing systems~\cite{vaishampayan2023procedural} and LLM-generated feedback in mock tech interview settings~\cite{swanand2025chase}. Research also offers solutions to mitigate bias and improve the decision-making capabilities of AI~\cite{eigner2024determinantsllmassisteddecisionmaking}. For instance, extensive work demonstrates the need for explainable~\cite{powell2024human}, transparent~\cite{vaishampayan2023procedural}, responsible~\cite{li2021algorithmic}, and accountable~\cite{yanamala2023transparency} AI systems in hiring contexts. To mitigate bias in AI and increase trust, research has explored a wide variety of techniques to promote responsible AI usage across domains. Techniques such as routine audits~\cite{yarger2020algorithmic}, ensemble learning with multiple LLM agents~\cite{radwan2024addressing}, and prompt engineering~\cite{kamruzzaman2024promptingtechniquesreducingsocial} have shown promise in mitigating bias and enhancing the accuracy of AI decisions. 

Further, the ``black box'' nature of AI-based systems in hiring contexts lead to significant lack of trust and desire for transparency among candidates~\cite{vaishampayan2023procedural}. This motivates the need for research tools demonstrate the effects of incorporating responsible and transparent AI practices in tech hiring pipelines. Governments have introduced laws to regulate and address the usage of AI in hiring pipelines (\ie recent laws on automated employment decision tools by the New York City (NYC) Department of Consumer and Worker Protection\footnote{\url{https://www.nyc.gov/site/dca/about/automated-employment-decision-tools.page}}), however these legal frameworks have ultimately failed~\cite{bust}. Thus, we posit research-based evidence is vital to continue highlighting the shortcomings of algorithmic hiring and demonstrate the effects of incorporating best practices for responsible AI in tech hiring pipelines.

\paragraph{Predictive Factors of SE Success}

A key goal of evidence-based hiring is to collect information from hiring processes to provide insights into factors predictive of job performance~\cite{tricarico2024evidence}. However, studies suggest current hiring processes fail to accurately predict the performance of candidates. For instance, sources suggest 75\% of companies report recent college graduate hires are unsatisfactory, with six our of 10 companies firing a recent hire within the last year~\cite{intelligent}. Evaluating software engineers in development contexts is a complicated problem. For instance, research posits a lack of effective metrics to assess the productivity of software engineers within development teams~\cite{sadowski2019rethinking}. To this end, research is needed to investigate metrics and techniques for predicting effective software engineers. Extensive research has explored predictive models in SE contexts,\footnote{\url{https://conf.researchr.org/home/promise-2025}} providing insights into factors predicting various aspects of software development---such as developer productivity~\cite{8643844}, accepted contributions to open-source repositories on GitHub~\cite{middleton2018contributions}, and defect locations within source code~\cite{turhan2009relative}. 

Future work can explore methods to predict the extent to which candidates can effectively contribute to software development teams. Hiring good programmers is critical for software development teams---however, hiring inadequate candidates is costly and can negatively impact development processes~\cite{HiddenCost}. Thus, methods to minimize false positive hires is critical for developing and maintaining high-quality software systems. Leveraging research-based evidence in hiring practices can support effective hiring decisions. For example, prior work outlines characteristics of 
``great'' software engineers---noting attributes across personal characteristics, decision-making abilities, technical expertise, and relationships with teammates~\cite{li2015makes}. Researchers can develop techniques to assess this information from tech hiring assessments and evaluate these traits in candidates based on the content of their resumes and tech interview performance.

\paragraph{Translating Research-Based Evidence to Practice}

A key challenge preventing research-based evidence in hiring pipelines is a lack of effective knowledge transfer, or the process of effectively conveying and applying research findings into practice~\cite{cartaxo2018towards}. The transfer of knowledge to practical contexts is critical in evidence-based decision making. For example, the field of translational medicine aims bridge the gap between clinical research and patient care~\cite{wehling2021principles}. However, in the SE domain, research highlights substantial challenges with sharing research findings to practitioners~\cite{winters2024thoughts,devanbu2016belief,wilson2024will}. To this end, novel solutions are needed to transfer evidence from research settings into tech hiring pipelines. Recently, tools have leveraged generative AI to support the translation of research findings into practice. For example, systems such as MirrorThink,\footnote{\url{https://mirrorthink.ai/}} Med-PaLM,\footnote{\url{https://sites.research.google/med-palm/}} and Rocket Matter,\footnote{\url{https://www.rocketmatter.com/chatgpt/}} leverage LLMs to synthesize research output for science, medical, and legal domains. Similar tools can be implemented to demonstrate the effectiveness of evidence-based strategies in SE contexts, such as tech hiring pipelines, to promote effective decision making for hiring software engineers.

\paragraph{Enhancing Evidence Effectiveness}

\begin{figure*}
    \centering
    \includegraphics[width=0.95\textwidth]{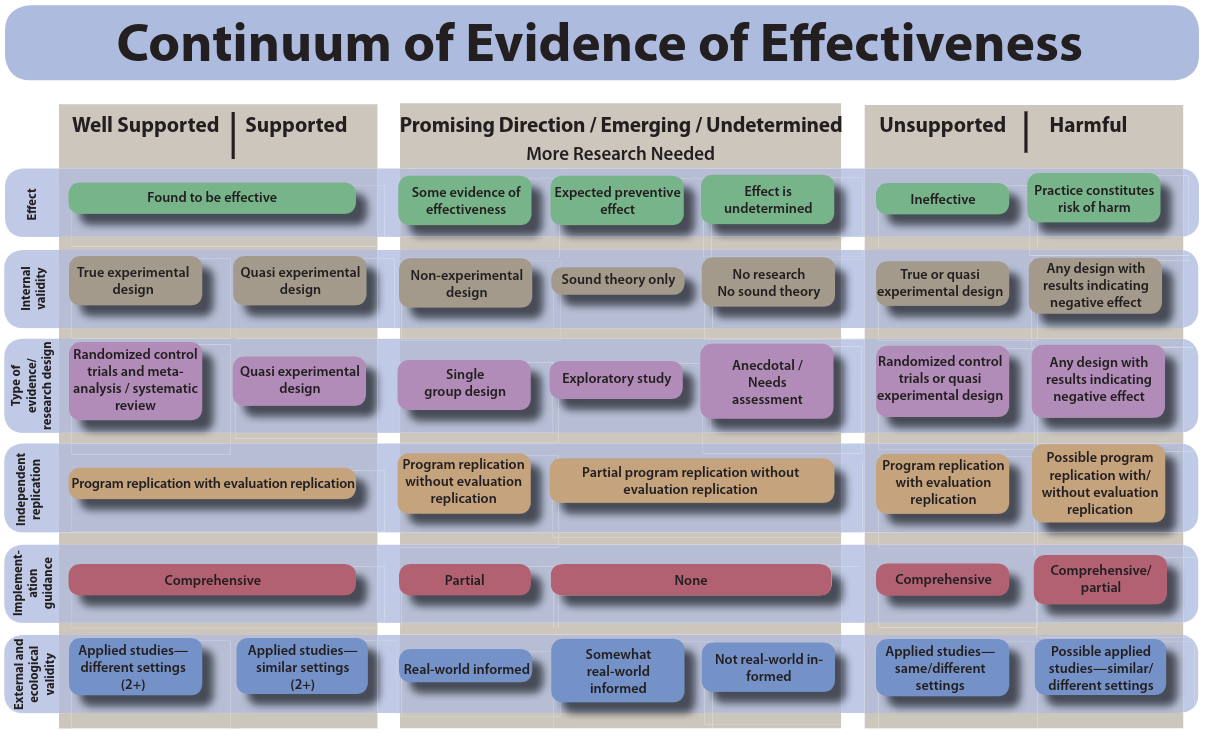}
    \caption{The Continuum of Evidence of Effectiveness Framework, as portrayed by~\cite{CDC2}.}
    \label{fig:continuum}
\end{figure*}

Evaluating the effectiveness of evidence is crucial for ensuring decision-making contexts are informed by the best available information. For instance, Brownson et al. outline review processes for evidence-based frameworks in public health~\cite{brownson1999evidence}. Similarly, Kitchenham et al. note that evaluating the effectiveness and efficiency of evidence-based software engineering is crucial for making incremental improvements~\cite{kitchenham2004evidence}. To assess evidence in decision-making contexts, the Center for Disease Control (CDC) offers the Continuum of Evidence of Effectiveness---a framework to evaluate various dimensions of evidence in various decision-making contexts across various dimensions (see Figure~\ref{fig:continuum})~\cite{CDC2}. The framework suggests more rigorous data collection and analysis methods promote stronger evidence to incorporate strategies into decision-making contexts~\cite{CDC-research}. For instance, to be considered as ``well-supported'' this framework suggests researchers conduct randomized control trials and systematic reviews, program replication, providing comprehensive replication guidelines, and applied studies in at least two different settings.

Further work is needed to investigate the effects of evidence on improving tech hiring decisions with regard to the six dimensions of the Continuum of Evidence of Effectiveness~\cite{CDC2,CDC-research}:
\begin{itemize}
    \item \textit{effect}, the effectiveness of strategy to impact the outcomes of interest---in this case, tech hiring decisions; 
    \item \textit{internal validity}, the extent to which short- and long-term outcomes can be attributed to the strategy; 
    \item \textit{research design}, the components of a research study (\ie measures, participant selection, etc.); 
    \item \textit{independent replication}, duplicating the strategy with a separate group of participants to determine if the same effects are achieved; 
    \item \textit{implementation guidance}, materials to aid in the implementation of strategies in different settings; and 
    \item \textit{external and ecological validity}, determining whether a strategy can demonstrate effects across a wide range of populations and contexts and in ``real life'' conditions.
\end{itemize}

Researchers suggest empirical SE research often suffers from a lack of rigor, lacking relevance to industry~\cite{winters2024thoughts} and standardized research methods~\cite{stol2018abc}. However, to date most research investigating tech hiring pipelines fails to address all six dimensions---lying within the unsupported to promising/emerging/undetermined effectiveness range. Researchers have offered numerous strategies to enhance tech hiring pipelines informed by real-world hiring problems, and provided evidence in support of the devised strategies (\ie~\cite{mahnaz2022async,swanand2025saner,behroozi2020does}). Yet, these efforts largely incorporate exploratory studies in irrelevant contexts with partial to no implementation guidance. For instance, our prior work introducing asynchronous technical interviews shows their effectiveness over traditional public interview settings for improving candidates' technical and communication abilities~\cite{mahnaz2022async}---however, our study consisted of a small effect ($n = 24$) on a limited set of LeetCode questions ($n = 5$) in one context without replication, in addition to lacking extensive implementation guidelines. Similarly, our work investigated GitHub-supported resume matching consists of formative and summative evaluations of a novel resume matching tool, but applies our approach to one setting in an exploratory user study design and limited implementation guidance and no replication. 

To enhance the effectiveness of research-based evidence, researchers should focus on addressing these dimensions in experiments---investigating them along with collecting perspectives from candidates and employers. This effort can enhance the quality of research evidence supporting strategies to improve decision-making in practical hiring contexts, motivating the transfer of knowledge and adoption of strategies into practice.

\section{Research Challenges and Opportunities}

We posit integrating contextual, experiential, and research-based evidence in tech hiring pipelines can enhance hiring decisions and form more effective development teams, improving the quality of software products. However, we also acknowledge challenges and new opportunities to incorporating evidence in hiring processes.

\subsection{Relevant Tech Hiring  Contexts}

One key challenge to promoting evidence-based decision making in tech hiring pipelines is investigating the proposed techniques in relevant contexts. Prior work suggests SE research in general suffers from an inability to conduct evaluations and apply findings to practical development contexts~\cite{winters2024thoughts}. Research exploring tech hiring pipelines faces the same limitation. For instance, prior work exploring tech hiring challenges uses synthetic resumes (\ie~\cite{skondras2023generating,glazko2024identifying}) and students conducting tech interviews in controlled settings (\ie~\cite{mahnaz2022async,behroozi2020does}). While these are effective for providing initial insights on hiring concepts, they fail to incorporate the high stakes nature of real-world hiring processes. Studies have also explored stakeholders' perceptions of various tech hiring modifications (\ie~\cite{mahnaz2022async,behroozi2020does,swanand2025chase,swanand2025saner})---however, we lack empirical insights on the effects of these interventions. For example, to effectively explore the effects of changes to hiring pipelines, researchers would need access to substantial amounts of data, including:

\begin{itemize}
    \item candidates' resume ranking/tech interview scores;
    \item the job description to which the candidate applied; 
    \item the outcome of the hiring process (\ie job offered or not); \item candidates' and employers' perceptions of the modified hiring process; and
    \item longitudinal data outlining candidates' contributions to the project after joining a development team.
\end{itemize}

Collecting this data would require close partnerships and common research agendas with industry collaborators~\cite{sjoberg2007future}---yet, these collaborations and participation from practitioners can be difficult to attain~\cite{10714522}. However, empirical evidence demonstrating the impact of incorporating evidence-based decision making in tech hiring pipelines, based on the Continuum of Evidence of Effectiveness framework~\cite{CDC2}, is necessary to promote evidence-based decisions and enact change to current hiring processes. 

\subsection{Toxic Tech Hiring Culture}

Another challenge is overcoming the overwhelming culture of tech hiring. For instance, despite the fact that research shows tech interviews are frustrating for candidates~\cite{mahnaz2019hiring}, insufficient for evaluating candidates' technical abilities~\cite{behroozi2020does}, and candidates praise companies that hire without whiteboards~\cite{HiringWithoutWhiteboards}---they remain widespread throughout the tech industry~\cite{whiteboard,CrackingCodingInterview}. Moreover, tech companies often pride themselves on having challenging hiring practices. For instance, Google has a high rate of false negatives in their hiring process, with a former senior staff engineer and manager stating:

\begin{quote}
``\textit{Google has a well-known false negative rate, which means we sometimes turn away qualified people, because that's considered better than sometimes hiring unqualified people. This is actually an industry-wide thing, but the dial gets turned differently at different companies. At Google the false-negative rate is pretty high. I don't know what it is, but I do know a lot of smart, qualified people who've not made it through our interviews. It's a bummer}''~\cite{google}.
\end{quote}

The current culture surrounding tech hiring has been described as ``toxic'' and ``awful''~\cite{awful}, requiring substantial time and effort to prepare~\cite{bell2023understanding} and invoking stress from participants~\cite{10.1145/3183399.3183415,hall2018effects}. For instance, studies analyzing comments from developers in online programming communities---such as Hacker News\footnote{A social network for software
practitioners at \url{https://news.ycombinator.com}} and Glassdoor\footnote{A job review website at \url{https://www.glassdoor.com}}---outline numerous frustrations and dissatisfaction of current tech hiring processes~\cite{mahnaz2019hiring,behroozi2020debugging}. Ultimately, the unnecessary complexities of tech hiring can cause candidates, who may be well-qualified for SE-related positions, to question their computing identity~\cite{lunn2021impact} and lose interest in tech careers~\cite{leaky}. To promote evidence-based practices in tech hiring pipelines, current perceptions and the culture surrounding tech hiring need to change.

\subsection{Attacks on Diversity}

The tech industry has a diversity problem, consisting of a predominantly white and male workforce while lacking representation of individuals from underrepresented gender identities, people of color, and other diverse backgrounds~\cite{lack}. This can lead to a lack of perspectives in software development processes, limiting innovation, quality, and development in software products~\cite{hyrynsalmi2024bridging,hyrynsalmi2024making}. Research shows current hiring processes can be biased. Studies show AI-based resume matching can be biased against underrepresented candidates (\ie~\cite{meyer2018amazon,glazko2024identifying,an2024large,an2024measuring}). For instance, Armstrong et al. show that, given identical resume content, LLMs provide higher ratings to resumes with White-sounding names compared to those corresponding to individuals from other racial and ethnic groups~\cite{armstrong2024silicone}. Further, prior work shows technical interviews are biased against individuals from minority backgrounds~\cite{hall2018effects,lunn2021impact}---favoring candidates with more time and resources (\ie money) to prepare~\cite{mahnaz2019hiring}. Research also shows enhancements to tech interview processes can improve the performance of underrepresented candidates. For example, Behroozi et al found that private~\cite{behroozi2020does} and asynchronous~\cite{mahnaz2022async} interview settings significantly improved the performance of candidates who identify as women. 

We posit that integrating evidence-based decision making into tech hiring processes can encourage more fair evaluations and promote a more diverse tech workforce. However, recent regulations by the United States government aims to diminish diversity, equity, and inclusion (DEI) efforts~\cite{eo}. As a result, many notable tech companies, such as Meta~\cite{meta}, Amazon~\cite{amazon}, and Google~\cite{google}, have eliminated DEI programs to recruit and support software engineers from minority backgrounds pursuing SE-related careers. Despite these trends, we posit evidence-based hiring practices can incur benefits to employers---reducing costs associated with tech hiring~\cite{HiddenCost} and ultimately providing more accurate assessments of candidates' abilities and fit within organizations. Future work is also needed to provide evidence demonstrating the effects and potential benefits of diverse software development teams in practical SE contexts~\cite{hyrynsalmi2024bridging}---highlighting the need for inclusive tech hiring practices.

\subsection{Evidence Beyond Hiring Pipelines}

This paper provides insights on using evidence to improve hiring decisions in the tech industry. However, support for software engineers is also critical after developers complete tech hiring pipelines and decide to join organizations. To this end, future work is needed to further collect contextual, experiential, and research-based evidence to support decision-making contexts in SE-related work environments. For example, Kitchenham et al. argue evidence-based practices in software engineering can enhance the development and maintenance of software products, offering insights to promote evidence-based software engineering based on medical domains~\cite{kitchenham2004evidence}. However, prior work shows developers often rely on their own beliefs instead of evidence~\cite{passos2011analyzing,devanbu2016belief}.

In addition to enhancing development practices, evidence is needed to further support candidates outside of hiring pipelines---as the overall tech pipeline remains leaky for software professionals who attained positions in tech industry~\cite{leaky}. For instance, as raises and promotions are commonly based on performance---providing accurate evidence of developers' contributions to projects is crucial for standardizing decisions. As an alternative example, a recent study from researchers at Stanford suggests 9.5\% of software engineers are ``ghost engineers''---essentially getting paid to do nothing~\cite{denisov2024predicting}.\footnote{\url{https://x.com/yegordb/status/1859290734257635439}} However, this work has been shown to lack evidence to support this claim~\cite{ghost}. Similarly, recent layoffs at Meta to cut 5\% of its workforce ($\approx 3,600$ jobs) were described as performance-based~\cite{blackmonday}---however, there is a lack of evidence on what constitutes a ``low performer'' and multiple individuals reported being laid off for non-performance-based reasons~\cite{leave}. Further, toxicity and
bias are rampant in the tech industry beyond hiring practices, which
can inhibit the performance of software engineers~\cite{awful}. Thus, novel techniques are needed to extract evidence outlining the capabilities and contributions of developers from development con-
texts, practitioners’ experiences, and empirical research.

\section{Conclusion}

To hire software engineers, tech hiring pipelines are often used to evaluate the technical proficiency of candidates across multiple phases---including resume matching and technical interviews. However, current tech hiring processes fail to accurately assess the skills of candidates and their ability to successfully contribute to development teams. To this end, we explore techniques to incorporate an evidence-based decision making framework~\cite{CDC2} into tech hiring pipelines---providing research directions to promote the collection of contextual, experiential, and research-based evidence in resume matching and technical interview assessments to inform hiring decisions. We also outline research challenges and opportunities to integrating these forms of evidence in tech hiring pipelines.

\bibliographystyle{ACM-Reference-Format}
\bibliography{main}
\end{document}

%% file: alias.tex
\newcommand{\ie}{\textit{i.e.,}\xspace}